\documentclass[a4paper,onecolumn,superscriptaddress,11pt,unpublished]{quantumarticle}
\pdfoutput=1
\usepackage[utf8]{inputenc}
\usepackage[english]{babel}
\usepackage[T1]{fontenc}
\usepackage{amsmath}
\usepackage{hyperref}

\usepackage{tikz}
\usepackage{lipsum}

\newcommand{\ket}[1]{\ensuremath{\left|#1\right\rangle}}
\newcommand{\upspin}{\ket{\uparrow}}
\newcommand{\downspin}{\ket{\downarrow}}

\begin{document}

\title{Twisted Spin in Quantum Mechanics}
\date{June 23, 2019}
\author{Stuart Samuel$^1$}

\email{stuartsamuel@hotmail.com}
\orcid{0000-0003-1830-0901}
\thanks{ \\
Retired Professor.  Previous institutions in historical order were: \\
\hspace*{1.5 mm} Institute for Advanced Study in Princeton (Member), \\
\hspace*{1.5 mm} Columbia University in New York (Professor of Physics), \\
\hspace*{1.5 mm} City University of New York  (Professor of Physics), \\
\hspace*{1.5 mm} Lawrence Berkeley National Laboratory (Scientist).
 }

\maketitle

\begin{abstract}
In quantum mechanics, it is often thought that the spin of an object points in a fixed direction at any point in time. 
For example, after selecting the $z$-direction as the axis of quantization, a spin-$1 \over 2$ object (such as an electron) may either point up or down. 
The spin can also be a linear combination of these two states, 
in which case, there is an axis in another direction in which the spin points in that direction. 
In this article, we focus on the spin-$1 \over 2$ case and point out that spin may not necessarily point in a fixed direction, 
a phenomenon that we call {\it twisted spin}. 
We argue that twisted spin occurs in nature, that at least some degree of twisting is generic, 
and propose an experiment to verify its existence.
\end{abstract}

\section{Introduction and Notation}
\label{intro}
Twisted spin can arise when an object of  non-zero spin passes through a magnetic field with a gradient. 
For simplicity, we take the object to have zero charge 
so that the only interaction with the magnetic field is through the magnetic moment of the object. 
In addition, in the region of interest, the gradient of the magnetic field is assumed to be constant. 
This relatively simple system can be analyzed through numerical integration. 
The results are presented in the next section. 
In section \ref{verificationExperiment}, we describe an experiment to measure twisted spin. 

There are many publications (a sampling of which are refs.\cite{Scullyetal,Diazetal,Porteletal,Hsuetal,Reddyetal}) 
providing the spatial dependence of the wavefunction of a particle with spin moving through a gradient magnetic field.  
Although countless articles discuss the procession of spin, that is, how spin varies {\it with time} in gradient magnetic fields,
we have been unable to find any publications that discuss the variation of spin {\it with space} at a fixed time. 
The purpose of this work is to create greater awareness of this phenomenon and 
for use in a future paper on how measurement works in quantum mechanics.\cite{SSSternGerlach}

In quantum mechanics, spin is quantized.\cite{Schiff} 
The mathematical procedure is to select an axis of quantization -- for example, the $z$ axis. 
When an object has spin $s = n/2$ for some integer $n$, 
there are $n+1$ eigenstates of the $z$-component of the spin operator $S_z$ with eignenvalues $-s, -s+1, \ldots, s-1, s$. 
The corresponding states are usually denoted by $\ket{s m_z}$, where $m_z$ is the eigenvalue of $S_z$. 
The most general spin state is a linear combination of these $n+1$ states. 
In this article, we focus on objects of spin-$1 \over 2$ and discuss the generalization to higher spins in the Conclusion. 
In this case, the notation $\upspin$ for $\ket{\frac{1}{2}, \frac{1}{2}}$ 
and $\downspin$  for $\ket{\frac{1}{2}, -\frac{1}{2}}$ is commonly used. 

\section{Solution of the Schr{\"o}dinger Equation in a Gradient Magnetic Field}
\label{mainbody}
\begin{figure}[h]
\centering
\centerline{\includegraphics[totalheight=0.30\textheight]{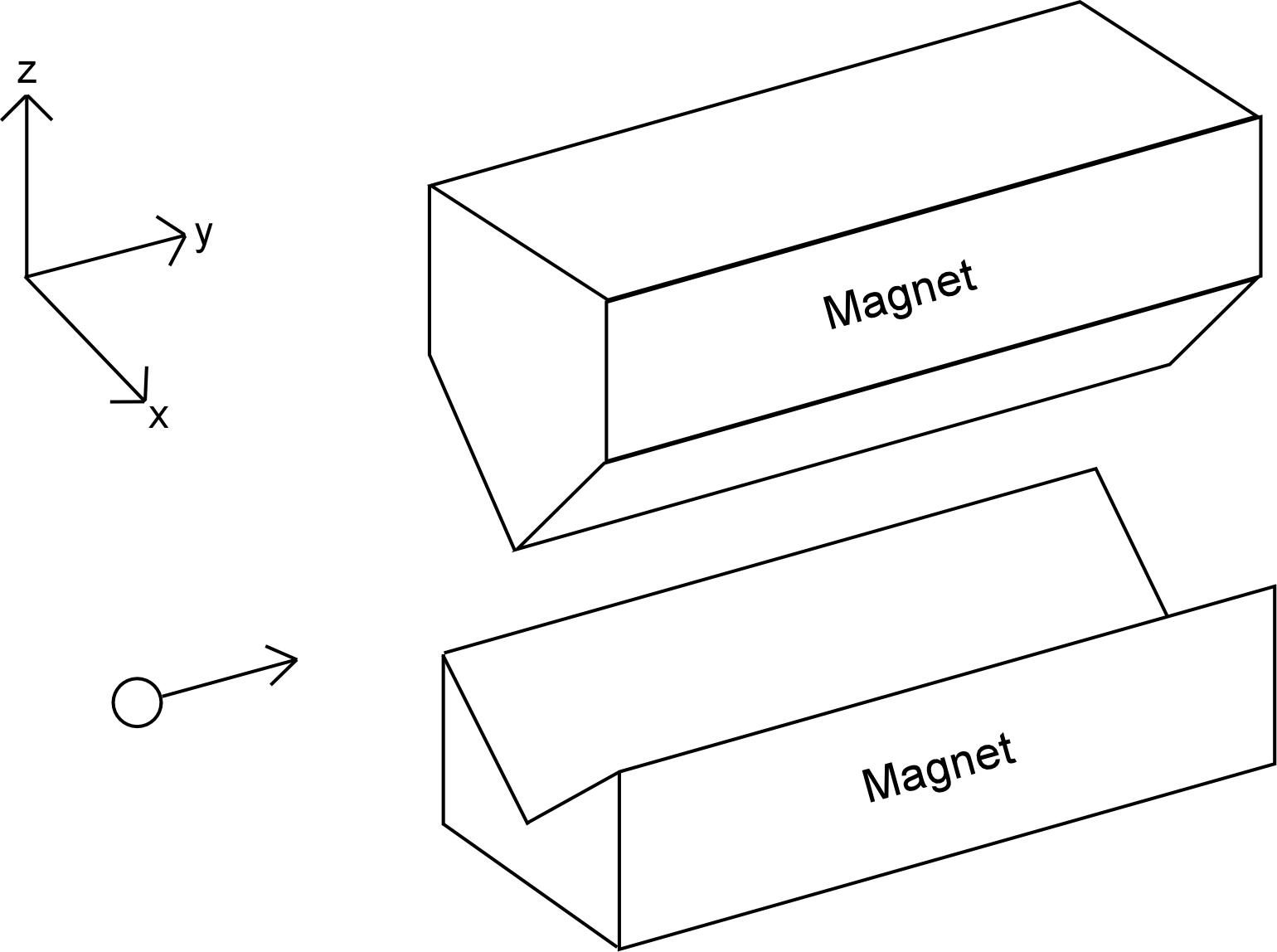}}
\vspace*{8pt}
\caption{The Spin-$1 \over 2$ Object O Heads into a Region with a Gradient Magnetic Field}
\label{fig:setup}
\end{figure}

Consider the situation of a neutral spin-$1 \over 2$ object moving through a magnetic field
 with a reasonably strong gradient, such as in the case of the Stern-Gerlach measurement of spin.\cite{SternGerlach} 
Orient the magnetic field in the $z$ direction 
and consider the case when the object is initially moving in the $y$-direction and the spin is pointing in the $x$-direction. 
We select the axis of spin quantization to be in the $z$-direction. 
Then, the initial spin part of the wavefunction is  $(\upspin +\downspin)/\sqrt{2}$. 
The setup is show in Figure  \ref{fig:setup}. 
The only difference between this setup and that of the Stern-Gerlach experiment is that there is no detection screen. 

The Schr{\"o}dinger equation for this situation is
\begin{equation}
i\hbar \frac{\partial\Psi ( \mathbf{r}, t) }{\partial t} = - \frac{\hbar^2}{2m}\nabla^2\Psi ( \mathbf{r}, t)  + V(\mathbf{r})\Psi ( \mathbf{r}, t) 
\  ,
\label{SchrodingerEquation}
\end{equation}
$$
   V(\mathbf{r}) =  - {\boldmath \mu} \cdot  \mathbf{B} ( \mathbf{r}) = - \gamma m_z \hbar B_z (z) 
\  ,
$$
where $m$ is the mass of the object, $\hbar$ is Planck's constant over $2\pi$, $V$ is the potential energy 
that arises from the interaction of the magnetic dipole moment {\boldmath$\mu$} of the object 
with the magnetic file $\mathbf{B}$, and $\gamma$ is the gyromagnetic ratio. 

Because the potential only depends on $z$, it is possible to factorize the wavefunction. 
Indeed, a solution to Eq.(\ref{SchrodingerEquation}) takes the following form: 
\begin{equation}
\Psi ( \mathbf{r}, t) = \Psi_x (x, t)  \Psi_y (y, t)  \Psi_z (z, t) \
  , 
\label{Factored}
\end{equation}
$$
   \Psi_z (z, t) = (\Psi_+ (z, t) \upspin + \Psi_-(z, t) \downspin) / \sqrt{2} 
\  ,
$$
where, in order to produce the situation in Figure \ref{fig:setup}, 
$ \Psi_y (y, t)$ is a wavepacket moving in the y-direction, 
and $\Psi_x (x, t)$ is a “non-moving” wavepacket in the sense that its mean momentum is zero in the $x$ direction. 
In order for Eq.(\ref{Factored}) to be a solution of the $3$-dimensional Schr{\"o}dinger equation in Eq.(\ref{SchrodingerEquation}),
the $x$ and $y$ wavefunctions must satisfy the free Schr{\"o}dinger equation in a single variable, that is,
\begin{equation}
   i\hbar \frac{\partial\Psi_x (x, t) }{\partial t} = - \frac{\hbar^2}{2m} \frac {\partial^2 \Psi_x (x, t)}{ \partial x^2 } 
\  ,
\label{GeneralSolution}
\end{equation}
$$
   i\hbar \frac{\partial\Psi_y (y, t) }{\partial t} = - \frac{\hbar^2}{2m} \frac {\partial^2 \Psi_y (y, t)}{ \partial y^2 } 
\  ,
$$
and $\Psi_+ (z, t)$ and $\Psi_-(z, t)$ must satisfy
\begin{equation}
   i\hbar \frac{\partial\Psi_+ (z, t) }{\partial t} = - \frac{\hbar^2}{2m} \frac {\partial^2 \Psi_+ (z, t)}{ \partial z^2 } - \frac{\hbar \gamma B_z}{2}  \Psi_+(z, t) 
\  ,
\label{zSchrodingerEquation}
\end{equation}
$$
   i\hbar \frac{\partial\Psi_- (z, t) }{\partial t} = - \frac{\hbar^2}{2m} \frac {\partial^2 \Psi_- (z, t)}{ \partial z^2 } + \frac{\hbar \gamma B_z}{2}  \Psi_-(z, t) 
\  .
$$

For the purposes of illustrating twisted spin, 
we assume that in the region of support of the wavefunctions, that $dB(z)/dz = B'_0$ is a constant and write $B_z = z B'_0$.\footnote{
Technically speaking such a magnetic field does not exist because it does not satisfy ${\bf \nabla} \cdot {\bf B} = 0$. 
However, the purpose of our calculations is to generate the differential $z$-dependence for up and down spin components
that is observed when spin-$1 \over 2$ objects enter a gradient magnetic field oriented in the $z$-direction. 
}
Then the potential becomes a linear function of $z$. 
As indicated in Eq.(\ref{zSchrodingerEquation}), the $z$-dependent part of the wavefunction separates into two motions: 
For the wavefunction component of spin up $\Psi_+ (z, t)$, that is, the coefficient of $\upspin$, the wave function will drift upward. 
For the $\downspin$ component $ \Psi_-(z, t)$, the wave functions will drift downward. 
This is the well-known effect of Stern-Gerlach of spatially separating up spin from down spin.

Choose units in which $\hbar = 1$. Then by scaling $z$ and $t$, Eq.(\ref{zSchrodingerEquation}) can be brought into the following form, 
which is useful for numerical integration: 
\begin{equation}
 i \frac{\partial\Psi_+ (z, t) }{\partial t} = - \frac{1}{2} \frac {\partial^2 \Psi_+ (z, t)}{ \partial z^2 } - 3 z  \Psi_+(z, t) 
\  ,
\label{ScaledSchrodingerEquation}
\end{equation}
$$
 i \frac{\partial\Psi_- (z, t) }{\partial t} = - \frac{1}{2} \frac {\partial^2 \Psi_- (z, t)}{ \partial z^2 } + 3 z  \Psi_-(z, t) 
\  .
$$

\begin{figure}[h]
\centerline{\includegraphics[ totalheight=0.30\textheight]{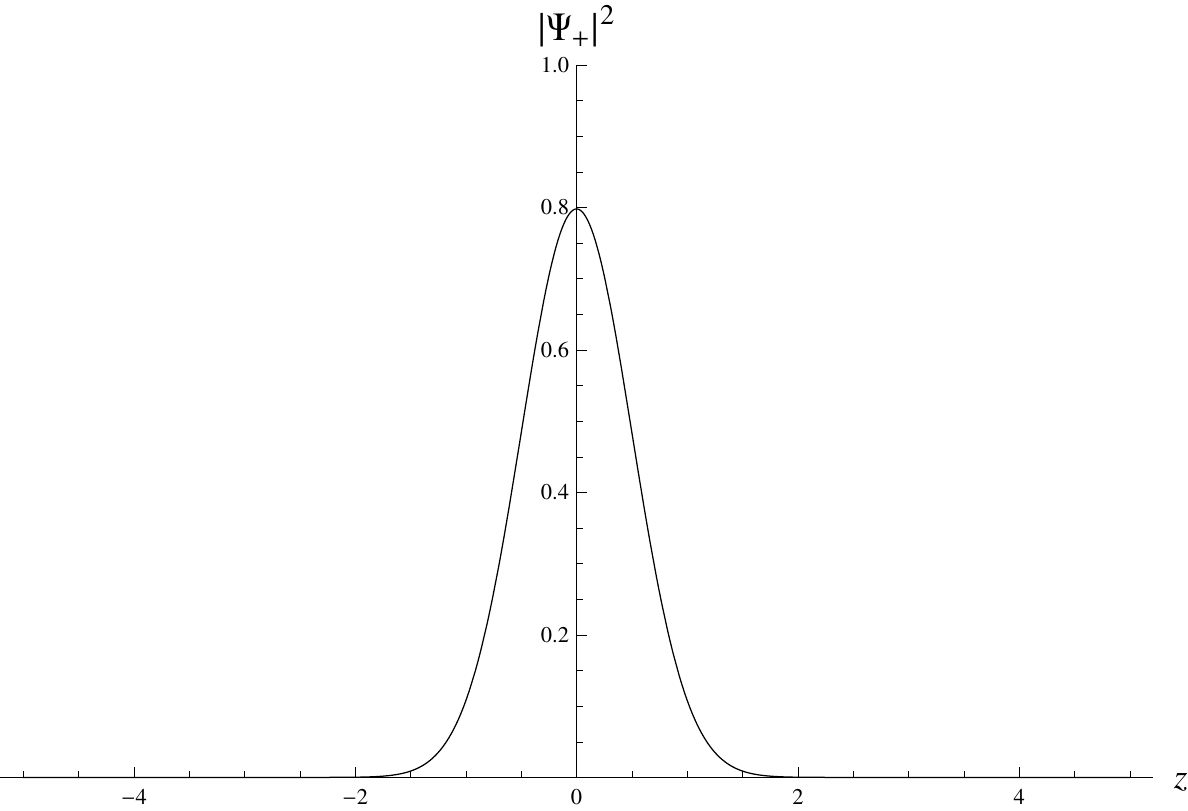}}
\vspace*{8pt}
\caption{The Initial Wavefunction Squared as a Function of $z$ at Time $=0.0$.}
\label{fig:InitialWaveFunctionSquared}
\end{figure}

The nature of the wavepackets packets $ \Psi_x (x, t)$ and $ \Psi_y (y, t)$ will not play a role in the analysis of twisted spin.
As for the $z$-dependence, we start with initial gaussian wavefunctions
\begin{equation}
 \Psi_+ (z, 0) = \Psi_- (z, 0) = \frac{ \exp ( - z^2  )  }{ (\pi / 2 )^{1/4} }
\  .
\label{InitialWavefunction}
\end{equation}
The probability of initially finding the particle at $z$, that is, the square of the wavefunction is displayed in Figure \ref{fig:InitialWaveFunctionSquared}. 
We numerically integrate Eq.(\ref{ScaledSchrodingerEquation}) subject to Eq.(\ref{InitialWavefunction}) and then examine the solutions at $t=1$. 
The Appendix shows the results for the wavefunctions graphically. 

Of interest is the direction of spin. 
If the spin is of the form 
$$
  (a +  i c) \upspin +  (b + i  d)\downspin 
\  ,
$$
where $a$, $b$, $c$ and $d$ are real numbers, then the spin points in the following direction:
\begin{equation}
  ( 2(a b + c d),  2( a d -b c) , a^2 - b^2 + c^2 - d^2) \frac{1}{ a^2 + b^2 + c^2 + d^2}  
\  .
\label{SpinDirection}
\end{equation}
If the wavefunction is normalized, then one can take $ a^2 + b^2 + c^2 + d^2$ to be $1$ in Eq.(\ref{SpinDirection}).

\begin{figure}[h]
\centerline{\includegraphics[trim=0cm 5cm 0cm 4cm, clip=true, totalheight=0.35\textheight]{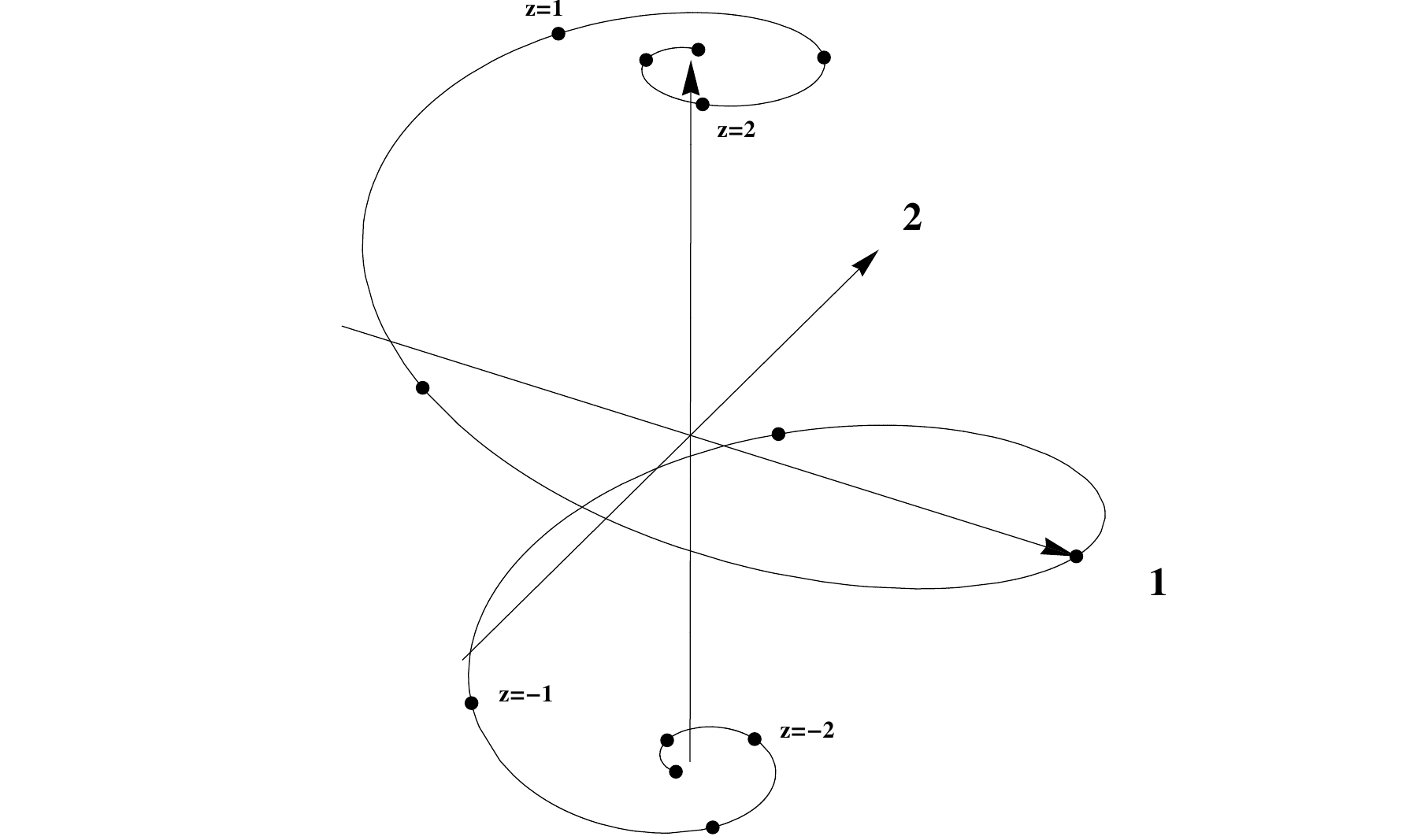}}
\vspace*{8pt}
\caption{Image of the Spin as a Function of $z$ at Time Equal to $1.0$.}
\label{fig:TwistedSpin}
\end{figure}

Because the coefficients of $\upspin$ and $\downspin$ in Eq.(\ref{Factored}) are different functions of $z$ for $t > 0$, 
the direction of the spin varies with $z$. 
For clarify, we use $1$, $2$ and $3$ as the labels for the directions $x$, $y$ and $z$ in spin space.
When $z$ is large and positive, 
the direction of the spin is very close to the positive $3$-axis and the spin is closed to $\upspin$. 
Similarly, when $z$ is large and negative, 
the direction of the spin is quite close the negative $3$-axis and the spin is closed to $\downspin$. 
When $z$ is of a modest value, the spin points in very different directions. 
Because $\Psi_+ (z, 1)$ and $\Psi_- (z, 1)$ are smooth functions of $z$, 
the spin will smoothly evolve from close to $\upspin$ for large postive $z$ 
to a direction close to $\downspin$ for large negative $z$. 
At $z=0$, the spin point in the positive $1$ direction, which is the original direction (at $t=0$).
Figure \ref{fig:TwistedSpin} on the next page shows the complete results for the spin at $t = 1$.
As one can see, the spin direction roughly traces out a helix. 
In short, due to the gradient magnetic field, the spin evolves with time and ends up not pointing in a single spatial direction.
It has become “twisted.” 

The particular form of the twisting depends on the initial state, the duration of time in the gradient magnetic field,  
and the strength of the magnetic field. 
Hence, Figure \ref{fig:TwistedSpin}, although exact at $t=1$ 
when Eqs.(\ref{ScaledSchrodingerEquation}) and (\ref{InitialWavefunction}) are used, 
is illustrative of how twisted spin occurs. 
Basically, whenever the up and down components of a spin-${1 \over 2}$ object develop
different spatial dependence, twisted spin arises 
as long as the component wavefunctions overlap 
and this is expected to happen in realistic situations. 

\section{A Verification Experiment}
\label{verificationExperiment}
\begin{figure}[!h]
\centerline{\includegraphics[ totalheight=0.30\textheight]{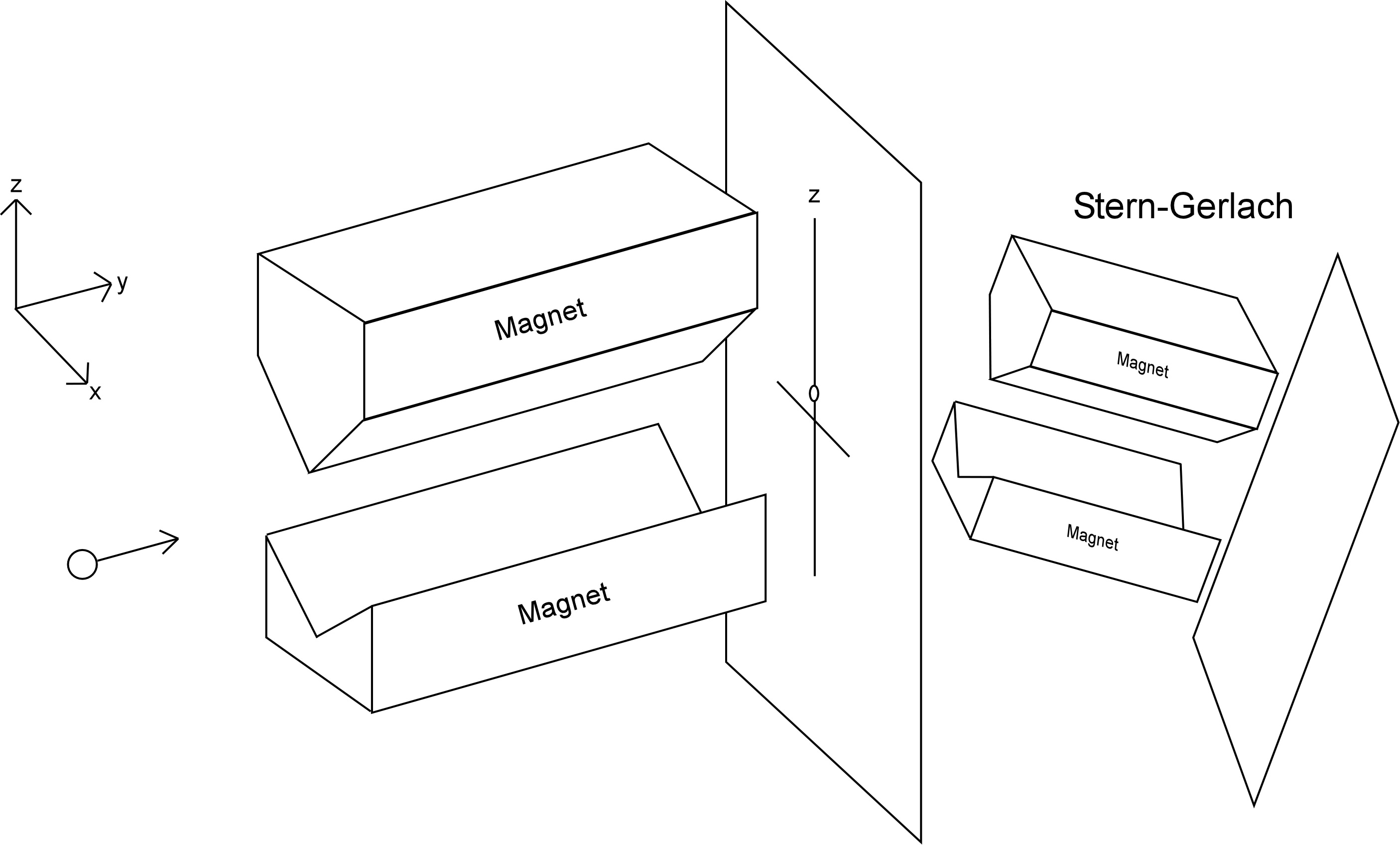}}
\vspace*{8pt}
\caption{An Experiment to Verify the Existence of Twisted Spin}
\label{fig:experiment}
\end{figure}

It is straightforward to devise an experiment to verify the existence of twisted spin. 
Place a screen behind Figure \ref{fig:setup} with a hole located on the $z$-axis. 
Behind the hole in the screen, put a spin-measuring apparatus (which can be a Stern-Gerlach device if oriented properly).
See Figure \ref{fig:experiment}. 
When spin-$1 \over 2$ objects with the same fixed initial spin are sent through the ``spin-twisting" gradient magnetic field, 
some will pass through the hole and their spin direction can be measured.
By locating the hole at various values of $z$ in the experimental setup of Figure \ref{fig:experiment}, 
one can verify that the spin is twisted. 

\section{Conclusions}
\label{conclusions}

We have show that, after an object of spin-$1 \over 2$ passes through a magnetic field with a non-zero gradient, 
its spin will vary as a function of position, and therefore there is no single direction in which the spin points 
at a given fixed time. 
This result holds as long as the spin and the magnetic field are not aligned. 
The result also holds for any object of non-zero spin: 
Objects of higher-than-$1 \over 2$ spin will have spin values that vary with position after passing through gradient magnetic fields. 
If the spin is $s$ then this takes place in a $4s+1$-dimensional space. 

Magnetic fields are omnipresent in the Universe. 
The flow of charged cosmic rays in space create weak magnetic fields. 
Both the Earth and Sun possess a magnetic field, and the same is undoubtedly true of other planets and stars. 
Among the strongest magnetic fields are those near neutron stars. 
None of these magnetic fields are uniform. 
Hence, an object of spin, whether it be near the surface of the Earth or in outer space, 
will eventually experience a magnetic field with a non-zero gradient and its spin will get twisted. 
The degree of twisting depends on the strength of the magnetic field gradients. 
In addition, twisting effects will tend to accumulate with time as the object passes through successive magnetic fields. 
If someday one is able to develop a device that can measure the twisted spin of an object with some degree of accuracy 
then such a device might be useful in assessing the past history of the object and the ambient magnetic fields 
through which it has passed. 

\bibliographystyle{plain}

\appendix

\section{Appendix: Wavefunctions at $t=1$}
\label{appendix}

\begin{figure}[h]
\centerline{\includegraphics[ totalheight=0.30\textheight]{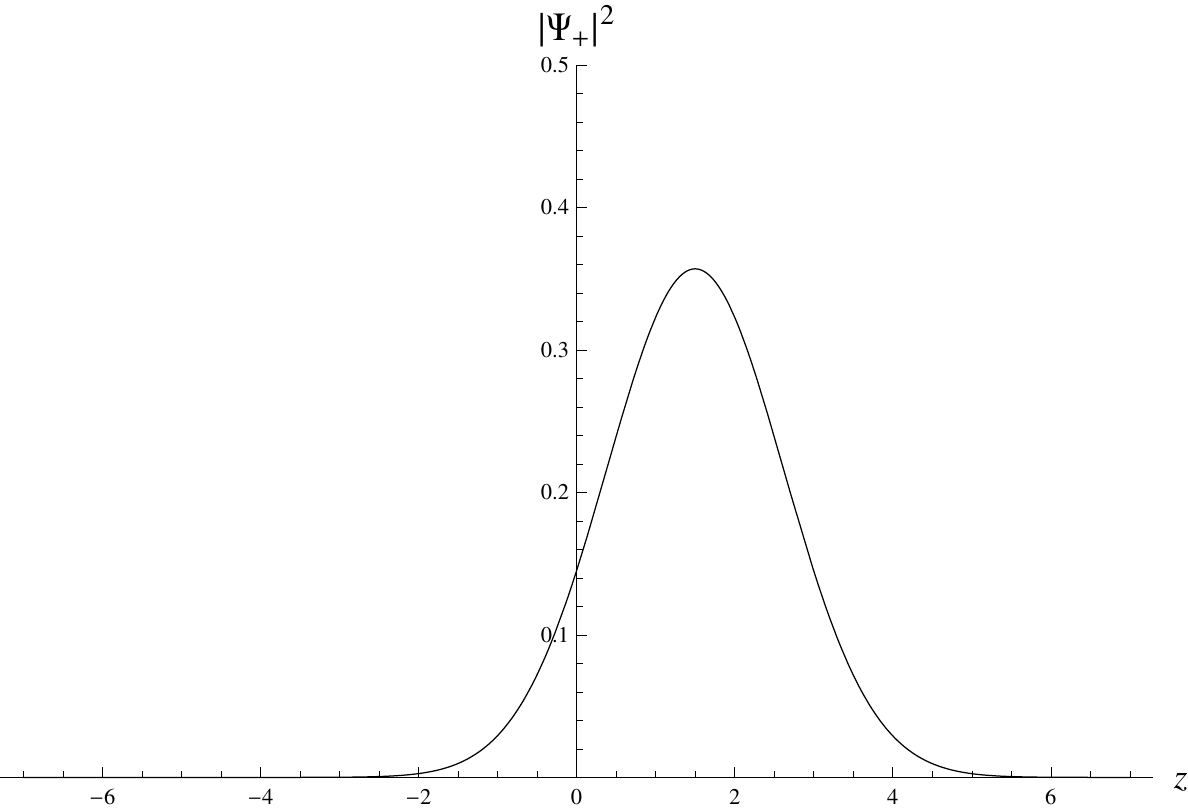}}
\vspace*{8pt}
\caption{The Square of the $\upspin$ Component of the Wavefunction of $z$ at Time Equal to $1.0$.}
\label{fig:PsiPlusSquared}
\end{figure}

\begin{figure}[h]
\centerline{\includegraphics[ totalheight=0.30\textheight]{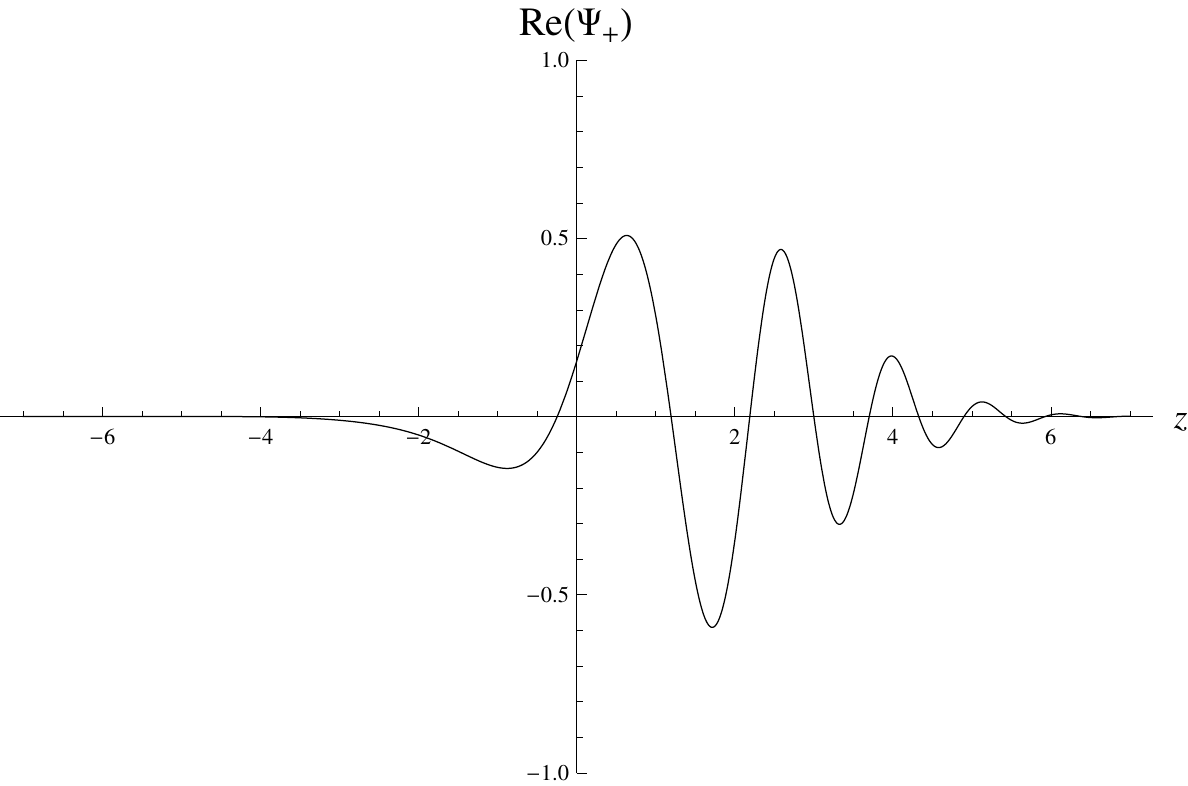}}
\vspace*{8pt}
\caption{The Real Part of the $\upspin$ Component of the Wavefunction of $z$ at Time Equal to $1.0$.}
\label{fig:RePsiPlus}
\end{figure}

\begin{figure}[t]
\centerline{\includegraphics[ totalheight=0.30\textheight]{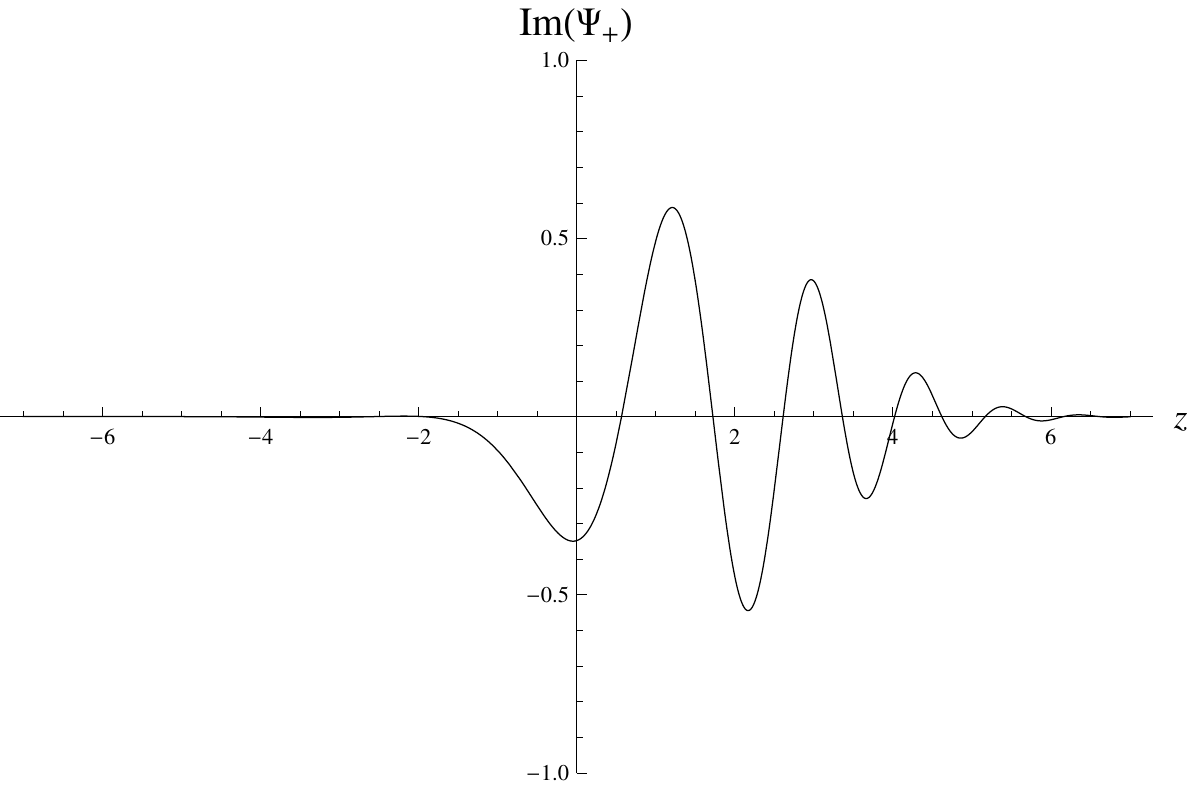}}
\vspace*{8pt}
\caption{The Imaginary Part of the $\upspin$ Component of the Wavefunction of $z$ at Time Equal to $1.0$.}
\label{fig:ImPsiPlus}
\end{figure}

When we integrate Eq.(\ref{ScaledSchrodingerEquation}) subject to the initial conditions in Eq.(\ref{InitialWavefunction}), 
we get the following results at $t=1$. 
For the $\upspin$ component of the wavefunction, namely, $\Psi_+$, 
the probability of finding the state at $z$ is show in Figure \ref{fig:PsiPlusSquared}.
As expected, 
the support of the wavefunction has shifted upward (the positive $z$ direction). 
The real and imaginary parts of  $\Psi_+$ are displayed in Figures \ref{fig:RePsiPlus} and \ref{fig:ImPsiPlus}.
The results for $\Psi_-$ are not graphically displayed here since they are ``mirror reflected": $\Psi_-(z) = \Psi_+(-z)$,
and the support of the $\Psi_-$ has shifted downward (the negative $z$ direction) as expected. 

\newpage
\clearpage

\ \\
\ \\
\ \\
\ \\
\ \\
\ \\
\ \\
\ \\
\ \\
\ \\
\ \\

\end{document}